\documentclass{ws-jaic}
\usepackage{ws-jaic-thm}        
\usepackage[square]{natbib}
\begin{document}

\markboth{Bensemann et al.}{Relating Blindsight and AI}

\catchline{0}{0}{0000}{}{}

\title{Relating Blindsight and AI: A Review}

\author{Joshua Bensemann\footnote{Corresponding author \\ josh.bensemann@auckland.ac.nz}, Qiming Bao, Gaël Gendron, Tim Hartill, Michael Witbrock}

\address{School of Computer Science, University of Auckland}

\maketitle

\pub{Received 3 May 2021}{Published: 9 July 2021}

\begin{abstract}
Processes occurring in brains, a.k.a. biological neural networks, can and have been modeled within artificial neural network architectures. Due to this, we have conducted a review of research on the phenomenon of blindsight in an attempt to generate ideas for artificial intelligence models. Blindsight can be considered as a diminished form of visual experience. If we assume that artificial networks have no form of visual experience, then deficits caused by blindsight give us insights into the processes occurring within visual experience that we can incorporate into artificial neural networks. This article has been structured into three parts. Section 2 is a review of blindsight research, looking specifically at the errors occurring during this condition compared to normal vision. Section 3 identifies overall patterns from Section 2 to generate insights for computational models of vision. Section 4 demonstrates the utility of examining biological research to inform artificial intelligence research by examining computation models of visual attention relevant to one of the insights generated in Section 3. The research covered in Section 4 shows that incorporating one of our insights into computational vision does benefit those models. Future research will be required to determine whether our other insights are as valuable.

\end{abstract}

\section{Introduction}	
A common assumption of Artificial Intelligence (AI) research is that AI agents currently lack the capacity for phenomenal experience \citep{block1995confusion}. Following that, any Deep Neural Network (DNN) trained to classify images does so without visual experiencing – AKA seeing – those images in the way humans do. Experiments have shown that DNN performance surpasses human accuracy in some visual tasks \citep{Eckersley2017}, yet those agents need to be retrained to obtained human-level performance in other visual tasks (see \cite{firestone2020performance} for a discussion on the fairness of these comparisons). While there are likely multiple reasons why human visual performance is currently considered superior, this review focuses on how visual experience impacts performance by examining those who have a reduced experiential capacity: blindsight patients.
\subsection{Definitions and History}
Visual experiences are part of phenomenal consciousness (P-Consciousness) which is related to yet separate from access consciousness (A-Consciousness) \citep{block1995confusion}. Information in a state of A-Consciousness can be manipulated by the intelligent agent whose "mind" the information resides – thoughts. This state occurs when the information is available for higher-order thought and can be reasoned over and used to determine the agent's actions, such as movement and speech. The critical difference between A-Consciousness and P-Consciousness is that the former is manipulating information without any experience of the information, whereas the latter is the perception of that information. For example, both humans and neural networks can make judgments using images, but only humans will visually experience (i.e., see) those images. Blindsight is an interesting case where information that would typically enter P-Consciousness (i.e., be seen) fails to do so, yet that information is still available for A-conscious processing.

The phenomenon of blindsight was initially observed in monkeys who had parts of brain area V1 (the primary visual cortex) removed but otherwise possessed normal eyesight.  These animals could still make some visual judgments, although their performance was impaired. Subsequent research has suggested that the primary cause of the ability to complete some visual task without V1 is due to multiple visual pathways within the brain; information would pass to subconscious areas of visual processing and allow the monkeys to complete specific tasks \citep{Stoerig1997}. The term blindsight was only used after similar observations were made in humans with damage to analogous areas of their brains. 

Since humans can self-report, researchers were able to determine that damage to V1 caused blind fields\footnote{the technical term for these blind fields is scotoma} - holes within their conscious vision \citep{Weiskrantz1996} . These blindsight patients are typically unable to have phenomenal experiences with any visual stimuli that enter these blind fields. However, there are some cases where specific manipulations to the stimuli can allow them to enter the patient's P-Consciousness. Researchers have found that some of these patients were still able to make simple visual judgments, despite being unaware that visual stimuli had been presented to them \citep{Weiskrantz1996}. Such findings are more common if the brain injury occurs earlier in life than later \citep{Fox2020}. These findings have driven many decades of research examining what happens during blindsight and visual processing when visual experience has been diminished. This article focuses only on research with humans to avoid assumptions about other animals' experiences. Those interested in animal results can find them in other reviewers of blindsight (e.g., \cite{Payne1996, Stoerig1997}).
\section{Blindsight Research }
\subsection{Location Tasks}
Early research with blindsight patients examined their accuracy at locating the source of visual stimuli \citep{Braak1971, Perenin1978a, Persaud2011, Poppel1973, WEISKRANTZ1974}. Such experiments typically used a horizontal array of lights where each light was tuned to project directly into the patient's blind field. Researchers would flash one light and then prompt the patient to locate its origin. The prompts, which were typically sounds, were required because the patient was unaware that the light had been presented. Experimenters found that patients' location accuracy was usually above chance levels, despite them being unaware of light.

The location task results also demonstrated that accuracy depended on which body part the patients used to locate the light. Some experiments required patients to move their eyes to the light's origin, and other experiments required patients to point their fingers at it. When locating by eye movements, the error margins were more considerable than when the patients were required to locate by pointing. Such results suggest that something about the different types of locating leads to noticeable differences in performance.

Experimental results also suggest interactions between accuracy and origin of a visual stimulus. Patients were most accurate at locating lights in the center of the array and became less accurate as the distance between the light and the center increased. Additionally, patients were more accurate when researchers increased either the flash duration or the projected stimuli' size \citep{Perenin1978a}.
\subsubsection{Motion Tasks}
Research has also tasked blindsight patients with making judgments about moving stimuli. In these experiments, lights were projected to the blind field and moved to another location within the field \citep{Barbur1980, Barbur1993, IsobelM.Blythe1986, Blythe1987}. Patients were then prompted to make a binary choice – typically left/right or up/down – about the light's direction.

Results from motion tasks show a relationship between the distance of movement and the patients' accuracy \citep{Barbur1980, IsobelM.Blythe1986, Blythe1987}. Patients became more accurate as the distance of movement increased. With that said, researchers kept the duration of the light's presentation constant, meaning that it is also true that accuracy increased as the velocity increased. It would be of interest to determine whether distance or velocity caused the increases, if not both. Additionally, experiments found that increasing the stimulus intensity (luminosity of the light) also increased the patients' accuracy \citep{Barbur1980}.

Another interesting test was group motion. If several objects move in random directions but are weighted to move in one direction overall, humans can determine the group's overall movement. Researchers have observed that while blindsight patients can determine one stimulus's movement, they fail at group movement tasks \citep{Azzopardi2001}.

Motion tasks have also demonstrated the effect of blindsight on the critical fusion frequency. Motion, especially in animation, depends on the ability to perceive several successive images as fluid motion. This fusion happens when the frequency of these images exceeds a minimum threshold – the critical fusion frequency. Tests have shown that this threshold was lower for stimuli presented to the blind field \citep{Barbur1980, Blythe1987}. In other words, an image that appears to be flashing on and off to a patent's normal visual field may appear to be on continuously when presented to that patient's blind field.
\subsection{Colour Discrimination Tasks}
Some tests have shown that blindsight patients can also discriminate between colours, despite being unable to perceive those colours \citep{Barbur1980, STOERIG1987, Stoerig1989, Stoerig1992, Weiskrantz1996, WEISKRANTZ1974}. Tests for color detection within blind fields have shown that discrimination curves are similar to those within normal visual fields. However, the general sensitivity of the blind fields is lower \citep{Barbur1980, Stoerig1989}. Lights with greater luminosity are required for the color to be detected.
\subsection{Object Discrimination Tasks}
Researchers have also shown that blindsight patients can discriminate between objects, with these discriminations ranging in difficulty. Tests have included making judgments about line orientations and discriminating between two shapes. Researchers have obtained mixed results; patients who perform accurately in one task can fail at others \citep{Barbur1980}.
Several line orientation tests have been done with blindsight patients. These procedures typically chose an orientation – for example, horizontal – and then present a line to the patient's blind field. Researchers then prompt the patient to determine whether that line was the same as the chosen orientation. Results show that while patients can perform this task, accuracy using the blind field is lower than of normal vision \citep{Barbur1980, WEISKRANTZ1987, WEISKRANTZ1974}.

More complicated tasks have had patients make judgments about object features. For example, one famous blindsight patient, DB, was required to identify X from O and was able to do so accurately \citep{WEISKRANTZ1987, WEISKRANTZ1974}. However, DB's accuracy at these tasks depended on the discrimination he was required to do. When asked to discriminate between triangles with straight edges versus those with curved, accuracy depended on how curved those lines were. Other tests included making square/rectangle judgments. DB's performance was relatively low in these tasks, although he was more became more accurate as the difference between long and short sides grew.

Researchers have also tested object discrimination and orientation simultaneously. Although patients could identify an equilateral triangle, they failed at determining whether it was pointing upwards or downwards \citep{Kentridge2015}.
\subsection{Affective Blindsight}
Researchers have also discovered that blindsight patients can make emotive judgments about stimuli presented to their blind field. This phenomenon, called affective blindsight, can also trigger responses from the emotional centers of the brain \citep{Burra2019, Celeghin2015} due to alternative visual pathways (see \cite{Gerbella2019}). Tests of affective Blindsight often involve identifying facial expressions; therefore, these tests are a type of object discrimination. 

It appears accuracy in affective blindsight depends on the type of stimulus that is processed. When researchers examined the effects of spatial frequency within an image, they found that low-frequency images lead to higher accuracy (defined by the level of activity in the amygdala) than high-frequency images \citep{Burra2019}.

Findings from affective blindsight studies also indicate a limit to what judgments patients can make from images of faces. For example, patients can judge simple emotions accurately, but not complex emotions such as arrogance or guilt \citep{Celeghin2017}. 
\subsection{Other Tasks and Results}
Having described many limitations of blindsight, we now turn to performance benefits. For example, when a famous patient, GY, performed a location task, researchers noted that although he had high accuracy with both blind fields and normal fields, GY's reaction times were faster during trials that presented stimuli to the blind field \citep{Persaud2011}. Being unaware of the stimuli lead to faster decisions that were also accurate.

Other benefits of blindsight include resistance to illusions. When reaching for an object, people typically shape their hand to an appropriate size for grabbing the desired object. This size adjustment occurs while their arm is still extending, i.e., before the hand gets to the object's location. Importantly, this adjustment is controlled by the object's perceived size; illusions that make the object appear larger or smaller cause people to adjust their hand to the incorrect size. When this test is performed blindsight patients, the hand adjustment is related to the actual size instead of the object's illusionary size (see \cite{Danckert2005}).
\section{Insights from Blindsight Research}
The purpose of this review was to identify how blindsight patients, who have diminished visual experience, perform visual tasks compared to those with normal vision. By doing so, we hoped to identify how differences between human and AI performance in visual tasks can be partially attributed to the function of visual experience.

The clear difference between visual performance in blindsight and normal vision is that those with normal vision make more accurate judgments than blindsight patients. Except for tasks involving illusions \citep{Danckert2005}, those with normal vision demonstrate superior performance in all areas. The research we have reviewed suggests that the lack of visual experience is correlated with impacts on performance. Of more interest to us is how and why these impacts affect accuracy.

Based on empirical results, some have suggested that blindsight patients' vision lacks the fine details present in normal vision that help us define the boundaries of an object \citep{WEISKRANTZ1974}. Being unable to determine when one object begins and ends makes it difficult to pinpoint its exact location, leading to increased error margins in location tasks. It also explains why patients perform poorly in tasks that used high-spatial-frequency stimuli \citep{Burra2019}. Having more details present in a facial image makes it harder to separate various components from each other and makes it harder to identify the facial expression. Similarly, being unable to determine the borders of a line would make it more challenging to identify the line properties such as orientation, curvature, and relative length, potentially explaining DB's performance in specific object discrimination tasks \citep{WEISKRANTZ1987}.

Related findings suggest that blindsight patients have an impairment in the physical detection of stimuli. Evidence for this is the replicated finding that increases in stimulus intensity or stimulus duration lead to increases in task accuracy. Both intensity and duration also increase accuracy in normal vision; however, the relative values required to achieve accurate performance in normal vision are much lower than those required for those with blindsight \citep{Barbur1980, Blythe1987}. Additionally, some patients can become aware of – but not necessarily identify – the stimulus if researchers increase that stimulus's intensity or duration beyond a threshold.

Alternatively, it may be more accurate to describe the above issues with stimulus selection rather than physical detection. Findings have shown that patients become aware of moving stimuli at lower luminosity values than stationary stimuli. Additionally, the luminosity required for detection decreases as the velocity increased. If blindsight is purely an impairment of detection, then the stimuli' velocity should be unrelated to whether the patient can see it. On the other hand, stimulus selection implies that something about the stimulus causes the visual system to attend to it and select it for further processing. This idea is also supported by results from neuroscience research showing that motion, luminosity, and duration affect the probability that a stimulus is attended \citep{Carrasco2011}. With all that said, both physical detection and stimulus selection are not mutually exclusive concepts; blindsight patients could have impairments in both.

Stimulus selection via visual attention is likely critical for explaining the performance gap between blindsight and normal vision. Attention is a core component in the Global Workspace Theory (GWT) \citep{Baars1988}. GWT explains conscious decision-making due to information entering a central workspace that broadcasts and receives information from multiple subconscious areas of the brain. A stimulus has to be attended to and selected for it to enter the global workspace. Once within the workspace, the stimulus is available for additional processing, increasing the accuracy of any judgments made about it. Impairments to stimulus selection imply that certain information is prevented from entering the workspace and, therefore, degrading performance in tasks involving that stimulus.

Inhibiting access to the global workspace also potentially explains failure at specific tasks and some of the few benefits of blindsight. Findings of patients failing at tasks such as identifying a triangle's orientation, detecting a group's overall motion, and determining complex emotions suggest that additional processing stages are missing in blindsight. Knowing the orientation of a triangle requires knowledge of which direction is up, determining overall movement requires the summation of individual movement, and determining emotions such as guilt requires knowledge of human social concepts. These tasks require the integration of information from multiple cognitive processes that are absent without the sharing of information; therefore, without first selecting the stimulus, these tasks become difficult, if possible at all. With that said, it does require time to share information; removing this restraint could lead to the increased reaction times observed in GY's performance \citep{Persaud2011}. 

Human brains have the advantage of starting with pre-learned information encoded into them via DNA, whereas most AI models are trained from scratch \citep{Zador2019}. While these genetic factors constrain the types of learning that the human can ultimately do, it also allows them to avoid learning foundational information that makes complex cognition possible. One such pre-learned function is visual attention \citep{Colombo2001}; humans already know what to attend to before they have to learn what they are attending during visual tasks. In contrast, visual attention mechanisms in AI models are typically learned while learning a primary task; the model determines what is worth attending based on what provides the most accurate method for determining the answer. While simultaneously training the attention mechanism and the primary decision-making process end-to-end does work, it can also lead to models making judgments based on attending the wrong thing (see \cite{Ribeiro2016}). However, AI vision models can benefit by pre-training to select and extract basic yet generally important visual information; this is related to the concept of training a brain set (see \cite{richards2019deep}). Of course, a claim of literal equivalence would require some definition of generally important, which may differ from what was important during the evolution of biological visual attention.

The actual difference between human and AI performance may be due to what happens to the information after it is selected or discarded. Following GWT, information selected via attention becomes available for additional processing. In DNNs, this additional processing occurs due to forward propagation to later levels in the network. In humans, this additional processing occurs due to broadcasting information back and forth from other specialized areas of the brain. In terms of AI models, this would be like training a multi-task model \citep{Caruana1997} and then having all of the separate tasks share their results – and possibly making additional judgments based on the shared results – before making a final decision. The main difference between what we suggest here and what has been done previously is that the model's other tasks may be seemingly unrelated to its primary task. The relevance of each task would emerge during the training of the final model.
\section{Relevant AI Models}
This review's central message is that superior human performance in visual tasks or AI systems is, in part, due to the experience of visual information. For this experience to occur, visual stimuli need to be selected during processing in the primary visual area via an attention process. Being attended does not guarantee that those stimuli are experienced, but experience does not occur unless those stimuli are attended. Due to the critical nature of attention within an experience, the final section examines attention mechanisms within deep learning research. 

There have been attempts to create DNNs with visual attention mechanisms (see \cite{Borji2013} for review). Here we have described some relevant research to show how effective this approach to computer vision has been. Perhaps future research will demonstrate that our other ideas help improve the current state of computer vision. 
\subsection{Deep Neural Network Attention Mechanisms}
\begin{figure}[!h]
\centering
\includegraphics[width=0.95\textwidth]{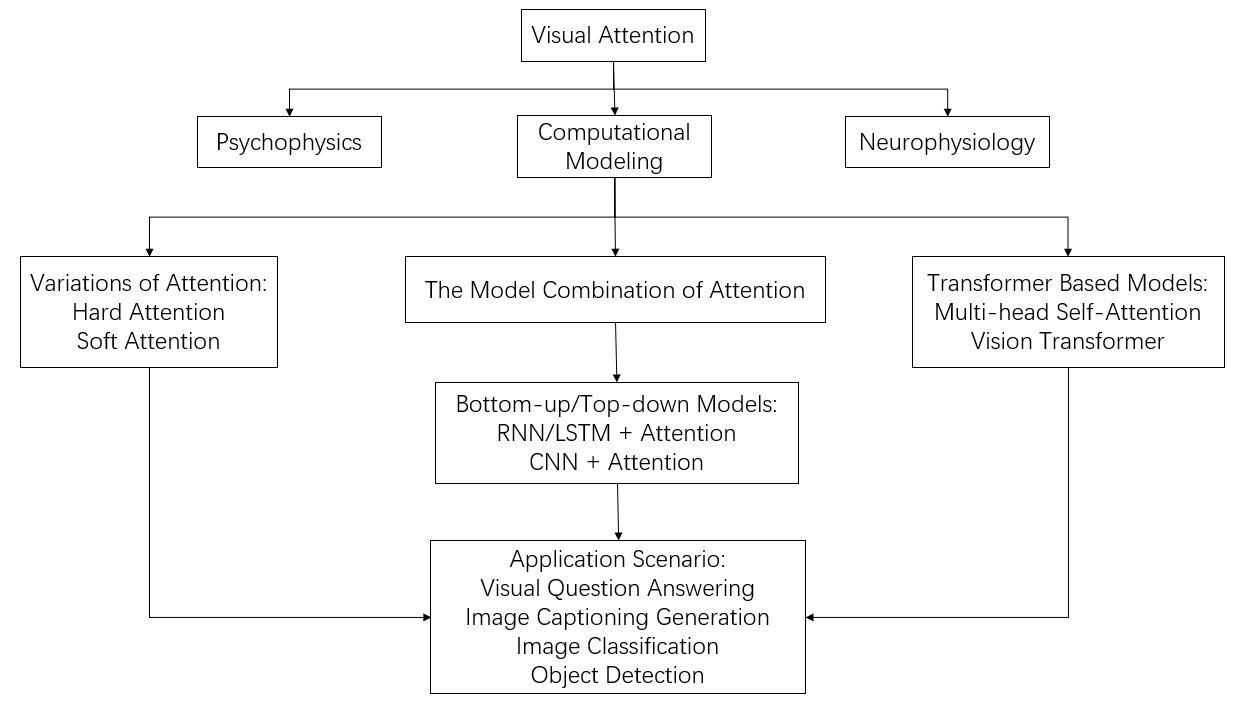}
\caption{A structure diagram of visual attention research.}
\label{attention}
\end{figure}

In this section, we have briefly introduced representative work involving visual attention mechanisms in DNNs. We can divide mainstream models using visual attention mechanisms into the three branches shown in Figure \ref{attention}. The first branch contains the two variants of attention mechanisms, hard and soft \citep{Xu2015}, where hard attention uses local calculation, and soft attention uses global calculation. In other words, hard attention methods focus on a small area of the visual field, and soft attention methods use the entire field. Soft attention is the more popular of the two methods, primarily because it is easier to train soft attention DNNs via backpropagation.

The second branch describes two attention methods, top-down and bottom-up. Two models that represent these methods are RNN/LSTM+Attention \citep{Anderson2018}, and CNN+Attention \citep{Xu2015}, respectively. The convolutional neural network (CNN), a revolutionary model also inspired by the human visual system, has an advantage in the image-level information extraction ability. CNNs can first extract the feature representation of the image, combine it with attention, and apply it to the Image Captioning Generation task. Similarly, RNN/LSTM can achieve good results on time-serialized data.


The third branch is the Transformer-based models that have achieved state-of-the-art (SOTA) performance in many Natural Language Process (NLP) tasks in the past two years, such as machine translation, question answering, reading comprehension, and several others. The major component of transformers is the unit of multi-head self-attention mechanism \citep{vaswani2017attention}. In the transformers encoder, the input is as a set of \textbf{key}-\textbf{value} pairs, \textbf{(K,V)}.  The dimensions of key and value are input sequence length, and both of them are the encoder hidden states. In the decoder, the previous output is represented as a query (\textbf{Q} of dimension $m$), and the following output is produced by mapping this query and the set of keys and values.
\begin{equation}\operatorname{Attention}(\mathbf{Q}, \mathbf{K}, \mathbf{V})=\operatorname{softmax}\left(\frac{\mathbf{Q K}^{\top}}{\sqrt{n}}\right) \mathbf{V}\end{equation}

The multi-head self-attention mechanism runs through the scaled dot-product attention multiple times in parallel. The independent attention outputs are concatenated and transformed into the dimensions of the target linearly.
\begin{equation}\begin{aligned}
\operatorname{MultiHead}(\mathbf{Q}, \mathbf{K}, \mathbf{V}) &=\left[\text { head }_{1} ; \ldots ; \text { head }_{h}\right] \mathbf{W}^{O} \\
\text { where head }_{i} &=\text { Attention }\left(\mathbf{Q} \mathbf{W}_{i}^{Q}, \mathbf{K} \mathbf{W}_{i}^{K}, \mathbf{V} \mathbf{W}_{i}^{V}\right)
\end{aligned}\end{equation}
where $W^{Q}_i$, $W^{K}_i$, $W^{V}_i$, and $W^O$ are parameter matrices to be learned.

There have been several attempts to apply transformer-based models to computer vision, including image classification, object detection, and visual question answering tasks, with great success \citep{Ramachandran2019, dosovitskiy2020image}. Such research has proved that the model relying only on Multi-head Self-Attention can have as good or better image feature extraction capabilities as typical CNN-based models.
\subsection{Graph Neural Network Attention Mechanisms}
A graph neural network is a class of DNNs where inputs take the form of graphs. These networks exploit the natural structure of these graphs to improve their performance over typical DNNs. Attention mechanisms have been used in conjunction with graph structures to improve visual models.

Understanding visual cues require an attention mechanism for focusing on the relevant visible elements of an image and the ability to interpret those elements in the context of the overall scene. A Scene Graph is a graph constructed from an input image that becomes a structured representation of that image. It represents the main objects visible on the image as nodes and their pairwise relationships with each other as edges. Its generation requires extracting not only features from the image but also relations. This problem involves visual attention for extracting the features and reasoning for establishing the corresponding relationships, thus understanding the image. Scene Graphs were first handcrafted and used for image retrieval \citep{johnson2015image} but its many following works focused on generating Scene Graphs \citep{xu2017scene, li2018factorizable, yang2018graph, zellers2018neural}. The critical idea shared by these works is the propagation of contextual information among the graphs' nodes. The use of Scene Graph allows for enhancing the detected features with high-level information. 

An application of Scene Graphs is for eXplanable Question Answering (XQA), and more particularly, Visual Question Answering (VQA). Humans can quickly process visual information for answering questions and retrospectively explain how they came to a conclusion. While Neural Networks are getting good at visual question answering, they are black-box systems that cannot explain how they answer a question. On the other hand, Scene Graphs offer the opportunity to return insights into how the system answered the question. \citep{ghosh2019generating} combine attention maps and Scene Graphs to produce explanations for QA.

Other propositions have been made for combining attention maps with graphs. The Factor Graph Attention model \citep{schwartz2019factor} adds textual information to the image; it combines image attention (attention map on an image) with a dialogue history and a question by a user about the image (how many people in the image? etcetera) in order to generate an answer to the question. The model is based on graph attention mechanisms, exchanging information between various sources, and focusing on multiple attention points. It is used in visual dialogue, a cognitive task created to give additional context for a task to imitate human-like decisions.

Structured Attention Graphs (SAGs) \citep{shitole2020structured} combines several attention maps to improve image classification. Each attention map focuses on a single region of the image and is combined with the others, becoming a node in a graph structure.

Co-saliency detection is another important visual task for a better understanding group of images or videos. It consists of detecting the common and distinct characteristics of images in a set. \citep{zhang2020adaptive} combines graph convolutional networks and attention graph clustering for extracting common and salient visual cues and discriminating the objects in the foreground from the background.
\subsection{Human Saliency Attention Mechanisms}
One final approach we will discuss is the use of saliency maps with neural networks. A saliency map is a representation of how human eyes shift their fixations across an image. Areas of the map are color-coded to represent the amount of time the eye spends in each image's location. As there is a direct link between attention and eye fixation, we can use saliency maps to determine which parts of an image draw the most attention by humans.

A large body of work has been directed at producing computational models that generate human saliency maps given an input of a particular image (for example \cite{li2016visual, ghariba2020novel, huang2015salicon, harel2007graph, tavakoli2017exploiting}). An approach for generating a visual saliency model based on Markov Chains \citep{harel2007graph} represents the image as a fully connected graph. Each pixel is a node of the graph, connected by an edge to its adjacent pixels. Attention weights are assigned to each node by running a Markov process on the graph. This achieves success in predicting human fixations points without requiring annotated data. Recent approaches \citep{li2016visual, ghariba2020novel, huang2015salicon, tavakoli2017exploiting} use convolutional neural networks in various configurations with training inputs being representative images and training labels being actual human saliency maps for these images that have been calculated by recording human eye fixation. Predicted saliency maps are typically represented as a matrix representing the field of vision, with elements being weights whose magnitude indicates the relative degree of attention a human might apply to that part of the image, or alternatively, the probability that a human might attend to that part.

Flores et al \citep{flores2019saliency} utilise such models (mainly \cite{tavakoli2017exploiting} and \cite{huang2015salicon}) to demonstrate improvement in object classification tasks. They use the models as pre-learned saliency map generators to create saliency maps for the images in their training set and then pair the resulting maps and images as dual input into a convolutional model that fuses the two inputs before running through further convolutional layers prior to classification. The saliency map generator plays a role conceptually similar to what genetically encoded information for attention in humans may play.

In a further series of experiments \citep{figueroa2020hallucinating}, it is demonstrated that such a pre-learned saliency model can be developed without specifically training to mimic human saliency. The maps subsequently generated enable almost as good performance on object classification tasks as the previously used human-imitating generators, indicating it is the "pre-learned" aspect rather than specifically the human aspect, which is of most value in this context. 

In this regard, the correspondence with improvements made in language understanding through pre-training and transfer learning in attention-based models such as BERT \citep{devlin2018bert} is rather striking. However, care must be taken in the comparison as these models utilize pre-trained word embeddings, pre-trained attention maps, and other model components. 
\section{Summary}
Blindsight is a fascinating phenomenon with implications for how the human visual system works. Decades of research have shown that although humans can process visual stimuli without awareness of those stimuli, task performance suffers without that awareness. It is generally assumed that current AI vision models lack awareness of the stimuli they process. Although no guarantee that creating models with awareness will dramatically improve their performance, it appears that when mechanisms important to awareness are incorporated into those models, they do improve performance.

\section*{Acknowledgments}
The authors would like to thank New Zealand's Tertiary Education Commission for providing grant money to establish the University of Auckland's Broad AI Lab. 

\backmatter

\bibliographystyle{ws-jaic}
\bibliography{references}

\begin{thebibliography}{57}
\newcommand{\enquote}[1]{#1}
\providecommand{\natexlab}[1]{#1}
\providecommand{\url}[1]{\texttt{#1}}
\providecommand{\urlprefix}{}
\expandafter\ifx\csname urlstyle\endcsname\relax
  \providecommand{\doi}[1]{doi:\discretionary{}{}{}#1}\else
  \providecommand{\doi}{doi:\discretionary{}{}{}\begingroup
  \urlstyle{rm}\Url}\fi

\bibitem[{Anderson \emph{et~al.}(2018)Anderson, He, Buehler \&
  Teney}]{Anderson2018}
Anderson, P., He, X., Buehler, C. and Teney, D. [2018] \enquote{{Bottom-Up and
  Top-Down Attention for Image Captioning and Visual Question Answering Visual
  attention},} \emph{Cvpr 20182} .

\bibitem[{Azzopardi \& Cowey(2001)}]{Azzopardi2001}
Azzopardi, P. and Cowey, A. [2001] \enquote{{Motion discrimination in
  cortically blind patients},} \emph{Brain} \textbf{124}(1),  30--46,
  \doi{10.1093/brain/124.1.30}.

\bibitem[{Baars(1988)}]{Baars1988}
Baars, B.~J. [1988] \emph{{A cognitive theory of consciousness}}, ISBN
  0521301335,
  \urlprefix\url{http://www.loc.gov/catdir/description/cam032/87020923.html}.

\bibitem[{Barbur \emph{et~al.}(1980)Barbur, Ruddock \& Waterfield}]{Barbur1980}
Barbur, J.~L., Ruddock, K.~H. and Waterfield, V.~A. [1980] \enquote{{Human
  visual responses in the absence of the geniculo-calcarine projection},}
  \emph{Brain} \textbf{103}(4),  905--928, \doi{10.1093/brain/103.4.905}.

\bibitem[{Barbur \emph{et~al.}(1993)Barbur, Watson, Frackowiak \&
  Zeki}]{Barbur1993}
Barbur, J.~L., Watson, J.~D., Frackowiak, R.~S. and Zeki, S. [1993]
  \enquote{{Conscious visual perception without VI},} \emph{Brain}
  \textbf{116}(6),  1293--1302, \doi{10.1093/brain/116.6.1293}.

\bibitem[{Block(1995)}]{block1995confusion}
Block, N. [1995] \enquote{On a confusion about a function of consciousness,}
  \emph{Behavioral and brain sciences} \textbf{18}(2),  227--247.

\bibitem[{Blythe \emph{et~al.}(1986)Blythe, Bromley, Kennard \&
  Ruddock}]{IsobelM.Blythe1986}
Blythe, I.~M., Bromley, J.~M., Kennard, C. and Ruddock, K. [1986]
  \enquote{Visual discrimination of target displacement remains after damage to
  the striate cortex in humans,} \emph{Nature} \textbf{320}(6063),  619--621.

\bibitem[{Blythe \emph{et~al.}(1987)Blythe, Kennard \& Ruddock}]{Blythe1987}
Blythe, I.~M., Kennard, C. and Ruddock, K.~H. [1987] \enquote{{Residual vision
  in patients with retrogeniculate lesions of the visual pathways},}
  \emph{Brain} \textbf{110}(4),  887--905, \doi{10.1093/brain/110.4.887}.

\bibitem[{Borji \& Itti(2013)}]{Borji2013}
Borji, A. and Itti, L. [2013] \enquote{{State-of-the-art in visual attention
  modeling},} \emph{IEEE Transactions on Pattern Analysis and Machine
  Intelligence} \textbf{35}(1),  185--207, \doi{10.1109/TPAMI.2012.89}.

\bibitem[{Braak \emph{et~al.}(1971)Braak, Schenk \& Vliet}]{Braak1971}
Braak, J. W. G.~T., Schenk, V. W.~D. and Vliet, A. G. M.~V. [1971]
  \enquote{{Visual reactions in a case of long-lasting cortical blindness},}
  \emph{Journal of Neurology, Neurosurgery {\&} Psychiatry} \textbf{34}(2),
  140--147, \doi{10.1136/jnnp.34.2.140}.

\bibitem[{Burra \emph{et~al.}(2019)Burra, Hervais-Adelman, Celeghin, de~Gelder
  \& Pegna}]{Burra2019}
Burra, N., Hervais-Adelman, A., Celeghin, A., de~Gelder, B. and Pegna, A.~J.
  [2019] \enquote{{Affective blindsight relies on low spatial frequencies},}
  \emph{Neuropsychologia} \textbf{128}(October 2017),  44--49,
  \doi{10.1016/j.neuropsychologia.2017.10.009}.

\bibitem[{Carrasco(2011)}]{Carrasco2011}
Carrasco, M. [2011] \enquote{{Visual attention : The past 25 years},}
  \emph{Vision Research} \textbf{51}(13),  1484--1525,
  \doi{10.1016/j.visres.2011.04.012},
  \urlprefix\url{http://dx.doi.org/10.1016/j.visres.2011.04.012}.

\bibitem[{Caruana(1997)}]{Caruana1997}
Caruana, R. [1997] \enquote{{Multitask Learning},} \emph{Machine Learning}
  \textbf{28},  46--47, \doi{10.1111/j.1468-0319.1995.tb00042.x}.

\bibitem[{Celeghin \emph{et~al.}(2015)Celeghin, de~Gelder \&
  Tamietto}]{Celeghin2015}
Celeghin, A., de~Gelder, B. and Tamietto, M. [2015] \enquote{{From affective
  blindsight to emotional consciousness},} \emph{Consciousness and Cognition}
  \textbf{36},  414--425, \doi{10.1016/j.concog.2015.05.007},
  \urlprefix\url{http://dx.doi.org/10.1016/j.concog.2015.05.007}.

\bibitem[{Celeghin \emph{et~al.}(2017)Celeghin, Diano, Bagnis, Viola \&
  Tamietto}]{Celeghin2017}
Celeghin, A., Diano, M., Bagnis, A., Viola, M. and Tamietto, M. [2017]
  \enquote{{Basic emotions in human neuroscience: Neuroimaging and beyond},}
  \emph{Frontiers in Psychology} \textbf{8}(AUG),  1--13,
  \doi{10.3389/fpsyg.2017.01432}.

\bibitem[{Colombo(2001)}]{Colombo2001}
Colombo, J. [2001] \enquote{{The Development of Visual Attention in Infancy},}
  \emph{Annual Review of Psychology} \textbf{52},  337--367.

\bibitem[{Danckert \& Rossetti(2005)}]{Danckert2005}
Danckert, J. and Rossetti, Y. [2005] \enquote{{Blindsight in action: What can
  the different sub-types of blindsight tell us about the control of visually
  guided actions?}} \emph{Neuroscience and Biobehavioral Reviews}
  \textbf{29}(7),  1035--1046, \doi{10.1016/j.neubiorev.2005.02.001}.

\bibitem[{Devlin \emph{et~al.}(2018)Devlin, Chang, Lee \&
  Toutanova}]{devlin2018bert}
Devlin, J., Chang, M.-W., Lee, K. and Toutanova, K. [2018] \enquote{Bert:
  Pre-training of deep bidirectional transformers for language understanding,}
  \emph{arXiv preprint arXiv:1810.04805} .

\bibitem[{Dosovitskiy \emph{et~al.}(2020)Dosovitskiy, Beyer, Kolesnikov,
  Weissenborn, Zhai, Unterthiner, Dehghani, Minderer, Heigold, Gelly
  \emph{et~al.}}]{dosovitskiy2020image}
Dosovitskiy, A., Beyer, L., Kolesnikov, A., Weissenborn, D., Zhai, X.,
  Unterthiner, T., Dehghani, M., Minderer, M., Heigold, G., Gelly, S.
  \emph{et~al.} [2020] \enquote{An image is worth 16x16 words: Transformers for
  image recognition at scale,} \emph{arXiv preprint arXiv:2010.11929} .

\bibitem[{Eckersley \& Nasser(2017)}]{Eckersley2017}
Eckersley, P. and Nasser, Y. [2017] \enquote{{EFF AI Progress Measurement
  Project},} \urlprefix\url{https://eff.org/ai/metrics}.

\bibitem[{Figueroa-Flores \emph{et~al.}(2020)Figueroa-Flores, Raducanu, Berga
  \& van~de Weijer}]{figueroa2020hallucinating}
Figueroa-Flores, C., Raducanu, B., Berga, D. and van~de Weijer, J. [2020]
  \enquote{Hallucinating saliency maps for fine-grained image classification
  for limited data domains,} \emph{arXiv preprint arXiv:2007.12562} .

\bibitem[{Firestone(2020)}]{firestone2020performance}
Firestone, C. [2020] \enquote{Performance vs. competence in human--machine
  comparisons,} \emph{Proceedings of the National Academy of Sciences}
  \textbf{117}(43),  26562--26571.

\bibitem[{Flores \emph{et~al.}(2019)Flores, Gonzalez-Garcia, van~de Weijer \&
  Raducanu}]{flores2019saliency}
Flores, C.~F., Gonzalez-Garcia, A., van~de Weijer, J. and Raducanu, B. [2019]
  \enquote{Saliency for fine-grained object recognition in domains with scarce
  training data,} \emph{Pattern Recognition} \textbf{94},  62--73.

\bibitem[{Fox \emph{et~al.}(2020)Fox, Goodale \& Bourne}]{Fox2020}
Fox, D.~M., Goodale, M.~A. and Bourne, J.~A. [2020] \enquote{{The Age-Dependent
  Neural Substrates of Blindsight},} \emph{Trends in Neurosciences}
  \textbf{43}(4),  242--252, \doi{10.1016/j.tins.2020.01.007},
  \urlprefix\url{https://doi.org/10.1016/j.tins.2020.01.007}.

\bibitem[{Gerbella \emph{et~al.}(2019)Gerbella, Caruana \&
  Rizzolatti}]{Gerbella2019}
Gerbella, M., Caruana, F. and Rizzolatti, G. [2019] \enquote{{Pathways for
  smiling, disgust and fear recognition in blindsight patients},}
  \emph{Neuropsychologia} \textbf{128}(August 2017),  6--13,
  \doi{10.1016/j.neuropsychologia.2017.08.028},
  \urlprefix\url{https://doi.org/10.1016/j.neuropsychologia.2017.08.028}.

\bibitem[{Ghariba \emph{et~al.}(2020)Ghariba, Shehata \&
  McGuire}]{ghariba2020novel}
Ghariba, B.~M., Shehata, M.~S. and McGuire, P. [2020] \enquote{A novel fully
  convolutional network for visual saliency prediction,} \emph{PeerJ Computer
  Science} \textbf{6},  e280.

\bibitem[{Ghosh \emph{et~al.}(2019)Ghosh, Burachas, Ray \&
  Ziskind}]{ghosh2019generating}
Ghosh, S., Burachas, G., Ray, A. and Ziskind, A. [2019] \enquote{Generating
  natural language explanations for visual question answering using scene
  graphs and visual attention,} \emph{arXiv preprint arXiv:1902.05715} .

\bibitem[{Harel \emph{et~al.}(2007)Harel, Koch \& Perona}]{harel2007graph}
Harel, J., Koch, C. and Perona, P. [2007] ``Graph-based visual saliency,'' in
  \emph{Advances in neural information processing systems}, pp. 545--552.

\bibitem[{Huang \emph{et~al.}(2015)Huang, Shen, Boix \&
  Zhao}]{huang2015salicon}
Huang, X., Shen, C., Boix, X. and Zhao, Q. [2015] ``Salicon: Reducing the
  semantic gap in saliency prediction by adapting deep neural networks,'' in
  \emph{Proceedings of the IEEE International Conference on Computer Vision},
  pp. 262--270.

\bibitem[{Johnson \emph{et~al.}(2015)Johnson, Krishna, Stark, Li, Shamma,
  Bernstein \& Fei-Fei}]{johnson2015image}
Johnson, J., Krishna, R., Stark, M., Li, L.-J., Shamma, D., Bernstein, M. and
  Fei-Fei, L. [2015] ``Image retrieval using scene graphs,'' in
  \emph{Proceedings of the IEEE conference on computer vision and pattern
  recognition}, pp. 3668--3678.

\bibitem[{Kentridge(2015)}]{Kentridge2015}
Kentridge, R.~W. [2015] \enquote{{What is it like to have type-2 blindsight?
  Drawing inferences from residual function in type-1 blindsight},}
  \emph{Consciousness and Cognition} \textbf{32},  41--44,
  \doi{10.1016/j.concog.2014.08.005},
  \urlprefix\url{http://dx.doi.org/10.1016/j.concog.2014.08.005}.

\bibitem[{Li \& Yu(2016)}]{li2016visual}
Li, G. and Yu, Y. [2016] \enquote{Visual saliency detection based on multiscale
  deep cnn features,} \emph{IEEE transactions on image processing}
  \textbf{25}(11),  5012--5024.

\bibitem[{Li \emph{et~al.}(2018)Li, Ouyang, Zhou, Shi, Zhang \&
  Wang}]{li2018factorizable}
Li, Y., Ouyang, W., Zhou, B., Shi, J., Zhang, C. and Wang, X. [2018]
  ``Factorizable net: an efficient subgraph-based framework for scene graph
  generation,'' in \emph{Proceedings of the European Conference on Computer
  Vision (ECCV)}, pp. 335--351.

\bibitem[{Payne \emph{et~al.}(1996)Payne, Lomber, Macneil \&
  Cornwell}]{Payne1996}
Payne, B.~R., Lomber, S.~G., Macneil, M.~A. and Cornwell, P. [1996]
  \enquote{{Evidence for greater sight in blindsight following damage of
  primary visual cortex early in life},} \emph{Neuropsychologia}
  \textbf{34}(8),  741--774, \doi{10.1016/0028-3932(95)00161-1}.

\bibitem[{Perenin \& Jeannerod(1978)}]{Perenin1978a}
Perenin, M.~T. and Jeannerod, M. [1978] \enquote{{Visual Function Within The
  Hemianopic Field Following Early Cerebral Hemidecortication In Man-I},}
  \emph{Neuropsychologia} \textbf{16}(1),  1--13.

\bibitem[{Persaud \emph{et~al.}(2011)Persaud, Davidson, Maniscalco, Mobbs,
  Passingham, Cowey \& Lau}]{Persaud2011}
Persaud, N., Davidson, M., Maniscalco, B., Mobbs, D., Passingham, R.~E., Cowey,
  A. and Lau, H. [2011] \enquote{{Awareness-related activity in prefrontal and
  parietal cortices in blindsight reflects more than superior visual
  performance},} \emph{NeuroImage} \textbf{58}(2),  605--611,
  \doi{10.1016/j.neuroimage.2011.06.081},
  \urlprefix\url{http://dx.doi.org/10.1016/j.neuroimage.2011.06.081}.

\bibitem[{Poppel \emph{et~al.}(1973)Poppel, Held \& Frost}]{Poppel1973}
Poppel, E., Held, R. and Frost, D. [1973] \enquote{{Residual Visual Function
  after Brain Wounds involving the Central Visual Pathways in Man},}
  \emph{Nature} \textbf{243},  295--296.

\bibitem[{Ramachandran \emph{et~al.}(2019)Ramachandran, Parmar, Vaswani, Bello,
  Levskaya \& Shlens}]{Ramachandran2019}
Ramachandran, P., Parmar, N., Vaswani, A., Bello, I., Levskaya, A. and Shlens,
  J. [2019] \enquote{{Stand-Alone Self-Attention in Vision Models},}
  (NeurIPS),  1--13, \urlprefix\url{http://arxiv.org/abs/1906.05909}.

\bibitem[{Ribeiro \emph{et~al.}(2016)Ribeiro, Singh \& Guestrin}]{Ribeiro2016}
Ribeiro, M.~T., Singh, S. and Guestrin, C. [2016] \enquote{{"Why should i trust
  you?" Explaining the predictions of any classifier},} \emph{Proceedings of
  the ACM SIGKDD International Conference on Knowledge Discovery and Data
  Mining} \textbf{13-17-Augu},  1135--1144, \doi{10.1145/2939672.2939778}.

\bibitem[{Richards \emph{et~al.}(2019)Richards, Lillicrap, Beaudoin, Bengio,
  Bogacz, Christensen, Clopath, Costa, de~Berker, Ganguli
  \emph{et~al.}}]{richards2019deep}
Richards, B.~A., Lillicrap, T.~P., Beaudoin, P., Bengio, Y., Bogacz, R.,
  Christensen, A., Clopath, C., Costa, R.~P., de~Berker, A., Ganguli, S.
  \emph{et~al.} [2019] \enquote{A deep learning framework for neuroscience,}
  \emph{Nature neuroscience} \textbf{22}(11),  1761--1770.

\bibitem[{Schwartz \emph{et~al.}(2019)Schwartz, Yu, Hazan \&
  Schwing}]{schwartz2019factor}
Schwartz, I., Yu, S., Hazan, T. and Schwing, A.~G. [2019] ``Factor graph
  attention,'' in \emph{Proceedings of the IEEE Conference on Computer Vision
  and Pattern Recognition}, pp. 2039--2048.

\bibitem[{Shitole \emph{et~al.}(2020)Shitole, Li, Kahng, Tadepalli \&
  Fern}]{shitole2020structured}
Shitole, V., Li, F., Kahng, M., Tadepalli, P. and Fern, A. [2020]
  \enquote{Structured attention graphs for understanding deep image
  classifications,} \emph{arXiv preprint arXiv:2011.06733} .

\bibitem[{Stoerig(1987)}]{STOERIG1987}
Stoerig, P. [1987] \enquote{{Chromaticity and Achromaticity},} \emph{Brain}
  \textbf{110}(4),  869--886, \doi{10.1093/brain/110.4.869}.

\bibitem[{Stoerig \& Cowey(1989)}]{Stoerig1989}
Stoerig, P. and Cowey, A. [1989] \enquote{{Wavelength sensitivity in
  blindsight},} \emph{Nature} \textbf{342}(6252),  916--918,
  \doi{10.1038/342916a0}.

\bibitem[{Stoerig \& Cowey(1992)}]{Stoerig1992}
Stoerig, P. and Cowey, A. [1992] \enquote{{Wavelength discrimination in
  blindsight},} \emph{Brain} \textbf{115}(2),  425--444,
  \doi{10.1093/brain/115.2.425}.

\bibitem[{Stoerig \& Cowey(1997)}]{Stoerig1997}
Stoerig, P. and Cowey, A. [1997] \enquote{{Blindsight in man and monkey},}
  \emph{Brain} \textbf{120}(3),  535--559, \doi{10.1093/brain/120.3.535}.

\bibitem[{Tavakoli \emph{et~al.}(2017)Tavakoli, Borji, Laaksonen \&
  Rahtu}]{tavakoli2017exploiting}
Tavakoli, H.~R., Borji, A., Laaksonen, J. and Rahtu, E. [2017]
  \enquote{Exploiting inter-image similarity and ensemble of extreme learners
  for fixation prediction using deep features,} \emph{Neurocomputing}
  \textbf{244},  10--18.

\bibitem[{Vaswani \emph{et~al.}(2017)Vaswani, Shazeer, Parmar, Uszkoreit,
  Jones, Gomez, Kaiser \& Polosukhin}]{vaswani2017attention}
Vaswani, A., Shazeer, N., Parmar, N., Uszkoreit, J., Jones, L., Gomez, A.~N.,
  Kaiser, L. and Polosukhin, I. [2017] \enquote{Attention is all you need,}
  \emph{arXiv preprint arXiv:1706.03762} .

\bibitem[{Weiskrantz(1987)}]{WEISKRANTZ1987}
Weiskrantz, L. [1987] \enquote{{Residual Vision in a Scotoma},} \emph{Brain}
  \textbf{110}(1),  77--92, \doi{10.1093/brain/110.1.77}.

\bibitem[{Weiskrantz(1996)}]{Weiskrantz1996}
Weiskrantz, L. [1996] \enquote{{Blindsight revisited},} \emph{Current Opinion
  in Neurobiology} \textbf{6}(2),  215--220,
  \doi{10.1016/S0959-4388(96)80075-4}.

\bibitem[{Weiskrantz \emph{et~al.}(1974)Weiskrantz, Warrington, Sanders \&
  Marshall}]{WEISKRANTZ1974}
Weiskrantz, L., Warrington, E.~K., Sanders, M. and Marshall, J. [1974]
  \enquote{Visual capacity in the hemianopic field following a restricted
  occipital ablation,} \emph{Brain} \textbf{97}(4),  709--728.

\bibitem[{Xu \emph{et~al.}(2017)Xu, Zhu, Choy \& Fei-Fei}]{xu2017scene}
Xu, D., Zhu, Y., Choy, C.~B. and Fei-Fei, L. [2017] ``Scene graph generation by
  iterative message passing,'' in \emph{Proceedings of the IEEE conference on
  computer vision and pattern recognition}, pp. 5410--5419.

\bibitem[{Xu \emph{et~al.}(2015)Xu, Ba, Kiros, Cho, Courville, Salakhutdinov,
  Zemel \& Bengio}]{Xu2015}
Xu, K., Ba, J.~L., Kiros, R., Cho, K., Courville, A., Salakhutdinov, R., Zemel,
  R.~S. and Bengio, Y. [2015] \enquote{{Show, Attend and Tell: Neural Image
  Caption Generation with Visual Attention},} \emph{Science of the Total
  Environment} \textbf{572},  169--176, \doi{10.1016/j.scitotenv.2016.07.196}.

\bibitem[{Yang \emph{et~al.}(2018)Yang, Lu, Lee, Batra \&
  Parikh}]{yang2018graph}
Yang, J., Lu, J., Lee, S., Batra, D. and Parikh, D. [2018] ``Graph r-cnn for
  scene graph generation,'' in \emph{Proceedings of the European conference on
  computer vision (ECCV)}, pp. 670--685.

\bibitem[{Zador(2019)}]{Zador2019}
Zador, A.~M. [2019] \enquote{{A critique of pure learning and what artificial
  neural networks can learn from animal brains},} \emph{Nature Communications}
  \textbf{10}(1), \doi{10.1038/s41467-019-11786-6},
  \urlprefix\url{http://dx.doi.org/10.1038/s41467-019-11786-6}.

\bibitem[{Zellers \emph{et~al.}(2018)Zellers, Yatskar, Thomson \&
  Choi}]{zellers2018neural}
Zellers, R., Yatskar, M., Thomson, S. and Choi, Y. [2018] ``Neural motifs:
  Scene graph parsing with global context,'' in \emph{Proceedings of the IEEE
  Conference on Computer Vision and Pattern Recognition}, pp. 5831--5840.

\bibitem[{Zhang \emph{et~al.}(2020)Zhang, Li, Shen, Liu, Chen \&
  Liu}]{zhang2020adaptive}
Zhang, K., Li, T., Shen, S., Liu, B., Chen, J. and Liu, Q. [2020] ``Adaptive
  graph convolutional network with attention graph clustering for co-saliency
  detection,'' in \emph{Proceedings of the IEEE/CVF Conference on Computer
  Vision and Pattern Recognition}, pp. 9050--9059.

\end{thebibliography}

\end{document}